\documentclass{INTERSPEECH2023}
\usepackage{multirow}
\usepackage{graphicx}
\usepackage[subtle]{savetrees}  
\usepackage{amsmath,graphicx,amssymb,delarray,subfigure,verbatim, bbding,boondox-calo}
\usepackage{diagbox, makecell}
\usepackage{bbding}
\usepackage{multirow}
\usepackage{color}
\usepackage{hyperref}

\interspeechcameraready

\title{Gesper: A Restoration-Enhancement Framework for General Speech Reconstruction}

\name{Wenzhe Liu$^{1}$, Yupeng Shi$^{1}$, Jun Chen$^{1,2}$, Wei Rao$^{1}$, Shulin He$^{1}$, Andong Li$^{3}$, Yannan Wang$^{1}$, Zhiyong Wu$^{2}$}

\address{
	 $^1$Tencent Ethereal Audio Lab, Tencent, Shenzhen, China \\
      $^2$Shenzhen International Graduate School, Tsinghua University, Shenzhen, China\\
      $^3$Institute of Acoustics, Chinese Academy of Sciences, Beijing, China 
}
\email{ \{wenzheliu, yupengshi, ellenwrao, ronslhe, yannanwang\}@tencent.com, y-chen21@mails.tsinghua.edu.cn, liandong@mail.ioa.ac.cn, zywu@se.cuhk.edu.hk}

\begin{document}

\maketitle

\begin{abstract}
This paper describes a real-time \textbf{Ge}neral \textbf{Spe}ech \textbf{R}econstruction (Gesper) system submitted to the ICASSP 2023 Speech Signal Improvement (SSI) Challenge.
This novel proposed system is a two-stage architecture, in which the speech restoration is performed, and then cascaded by speech enhancement.
We propose a complex spectral mapping-based generative adversarial network (CSM-GAN) as the speech restoration module for the first time.
For noise suppression and dereverberation, the enhancement module is performed with fullband-wideband parallel processing.
On the blind test set of ICASSP 2023 SSI Challenge, the proposed Gesper system, which satisfies the real-time condition, achieves 3.27 P.804 overall mean opinion score (MOS) and 3.35 P.835 overall MOS, ranked 1st in both track 1 and track 2.
\end{abstract}
\noindent\textbf{Index Terms}: speech signal improvement, two-stage, speech restoration, speech enhancement

\section{Introduction}
Real-time communication (RTC) systems such as teleconferencing systems, smartphones and telephones, have become a necessity in the life and work of individuals. In order to achieve high-quality communication experiences, it is crucial to address the challenges of speech signal quality in RTC systems. However, due to the influence of acoustical capturing, noise/reverberation corruption and network congestion, the speech quality of current RTC systems is still deficient.
The ICASSP 2023 SSI Challenge\footnote{\href{https://www.microsoft.com/en-us/research/academic-program/speech-signal-improvement-challenge-icassp-2023/}{https://www.microsoft.com/en-us/research/academic-program/speech-signal-improvement-challenge-icassp-2023/}} focuses on improving the speech signal quality in RTC systems, which involves tackling the difficulties of noise, coloration, discontinuity, loudness, and reverberation of speech in a variety of complex acoustic conditions. Noisiness includes background noise, circuit noise and coding noise. Coloration results from bandwidth limitation and frequency response distortions of the microphone. Packet loss results in speech discontinuity. Loudness problem includes clipping, nonlinear distortion and far-field recording.  

In the speech enhancement field, a ``noise suppression and speech restoration'' architecture has been proposed recently~{\cite{hao2020masking, liu2021know, du2020joint}}. These methods usually contain two-stage processing modules. In the first stage, The noise suppression (NS) module is used to reduce the noise or background components. However, the mask-based or mapping-based NS modules often adversely affect the speech component as more noise is suppressed, which tends to be increasing distortions of the speech or signal component. To reduce the speech degradation, the second stage module is adopted to re-process the NS enhanced speech and restore the higher-quality speech spectrum based on the time-frequency context information. Leaning on the strong generative capability of the vocoders, generative models have been introduced in the restoration stage~{\cite{maiti2020speaker, du2020joint, andreev2022hifi++}}. a vocoder such as WaveNet~{\cite{oord2016wavenet}}, LPCNet~{\cite{valin2019lpcnet} or WaveGlow~{\cite{prenger2019waveglow}} is applied in the restoration stage to re-generate speech waveform based on the Mel spectrum enhanced by the NS modules in the first stage. More recently, VoiceFixer~{\cite{liu2021voicefixer}} is also proposed to performed in the above-mentioned enhancement and restoration procedure including the noise reduction, dereverberation, bandwidth extension and clipping tasks.

However, due to the complexity of the acoustic scenarios provided in SSI Challenge, speech is distorted heavily. The above framework may further damage the quality of the speech. Excessive suppression of the degraded speech signal caused by the noise reduction methods may significantly increase the difficulty in restoring the desired speech signal without the guidance of semantic information. Therefore, a ``restoration and enhancement'' two-stage framework namely Gesper addresses the complicated problems in the SSI Challenge. Since the generation model in the time domain has poor high-frequency representation ability and abandons phase information, to overcome this limitation, a complex spectrum mapping-based generative model has been introduced. We first employ CSM-GAN as the restoration module for speech distortion restoration, narrowband bandwidth expansion (BWE) as well as preliminary denoising and dereverberation. Moreover, since there may still exist residual noise components and artifacts in the output of the restoration module, to further improve the quality of the speech signal, the enhancement module is applied in the second stage. As mentioned in \cite{zhang2022multi}, independent processing with wideband and fullband signal respectively improves speech enhancement performance which reduces the dynamic range of the spectrum. Parallel processing has been utilized to improve the efficiency of full-band speech enhancement. 

In summary, this paper makes the following contributions:
\begin{itemize}
    \item We propose a novel restoration-enhancement framework for general speech quality improvement, to address the difficulties of noise, coloration, discontinuity, loudness, and reverberation of speech.
    \item We design a complex spectrum mapping-based generation model for speech restoration, which shows better performance with respect to the previous vocoders. 
    \item We introduce a wideband and fullband parallel processing method for full-band speech enhancement.
\end{itemize}

The rest of the paper is organized as follows. In Section~{\ref{system-overview}}, we illustrate the overall diagram. We present the experimental setup in Section~{\ref{experiments}}. The results and analyses are given in Section~{\ref{results-and-analysis}}, and the final conclusions are drawn in Section~{\ref{conclusion}}.

\section{Proposed system}
\label{system-overview}
\label{overview}
\begin{figure*}[!htbp]
	\centering
    \includegraphics[width=0.85\linewidth]{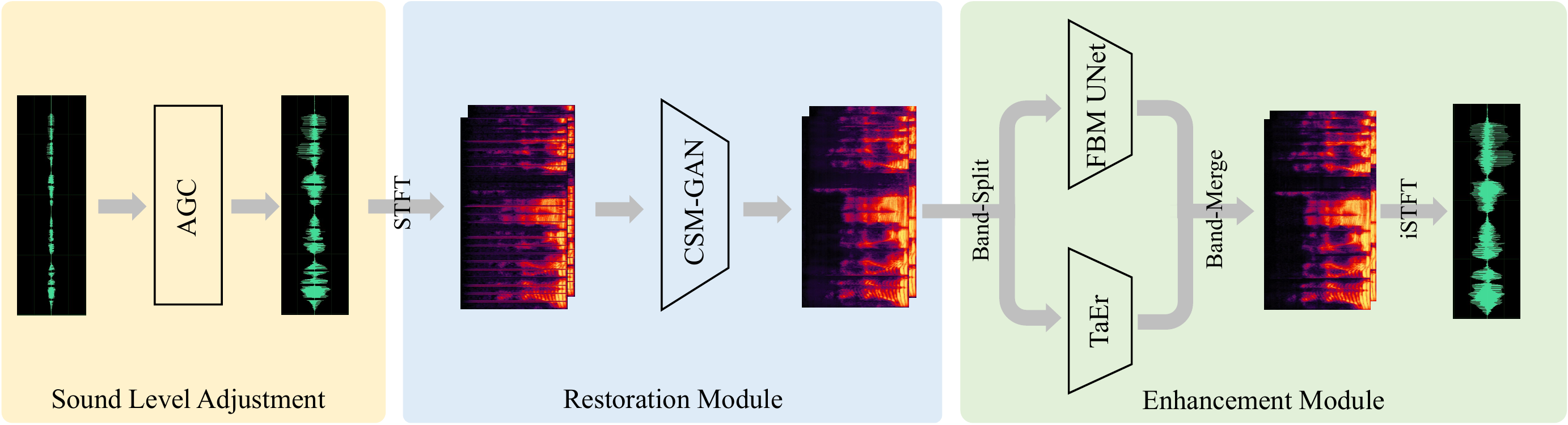}
	\caption{The general schematic of the proposed system. The "AGC" denotes auto gain control.}

	\label{fig:totalarch}
\end{figure*}

\begin{figure}[!htbp]
	\centering
    \includegraphics[width=1\linewidth]{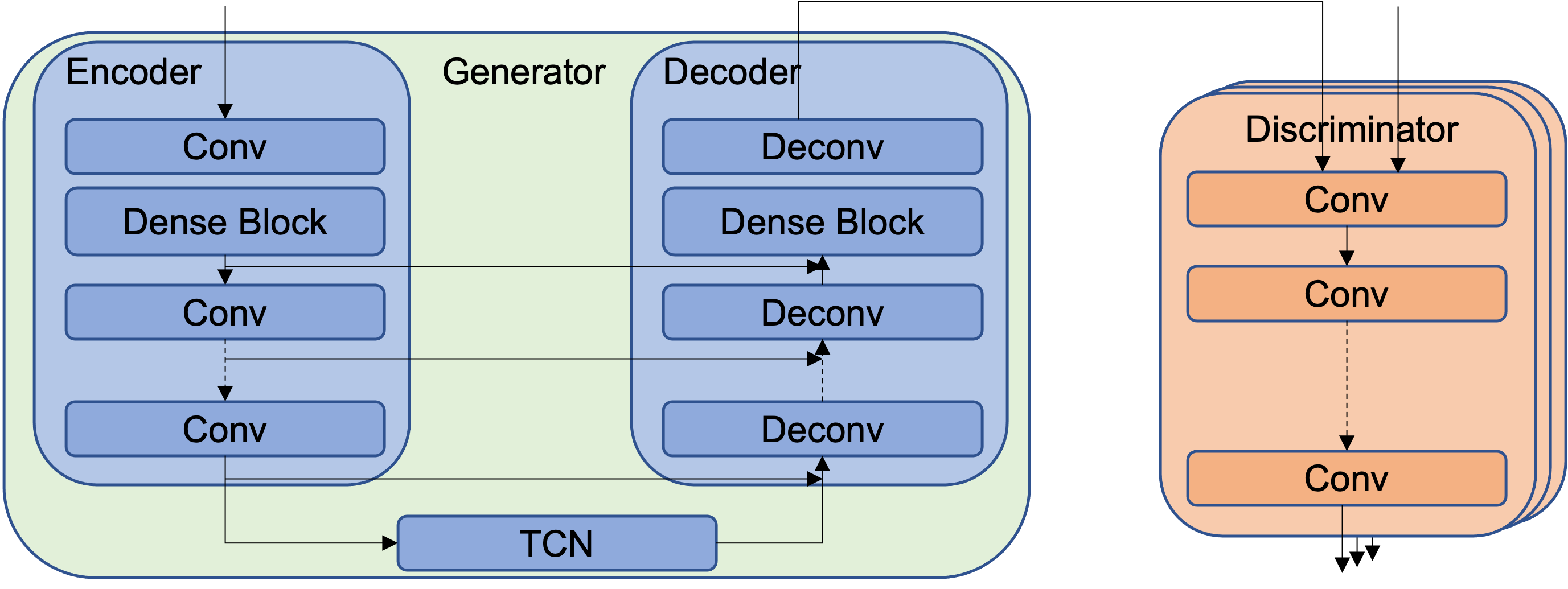}
	\caption{The architecture of CSM-GAN.}

	\label{fig:ganarch}
    \vspace{-0.3cm}
\end{figure}
As shown in Fig.\ref{fig:totalarch}, our proposed system is composed of three parts: sound level adjustment, restoration module, and enhancement module.
The sound level adjustment is adopted as data pre-processing and the restoration module and the enhancement module constitute the two-stage improvement framework.
The input audio waveform 
is first adapted to the appropriate volume by sound level adjustment, and then the short-time Fourier transform (STFT) is applied to obtain the complex spectrogram.
The real and imaginary parts of the complex spectrum are then fed into a two-stage architecture:
1) the restoration module first performs speech distortion restoration, and preliminary denoising and dereverberation with a generative adversarial network;
2) the enhancement module further eliminates residual noise components and artifacts based on a relatively high-quality speech complex spectrum generated by the restoration module.
Eventually, the output of the enhancement module passes through the inverse STFT (iSTFT) to yield the final prediction of the model.
Each module will be described in detail in the following parts.

\subsection{Sound Level adjustment}
\label{sec:Sound Level Adjustment}
The role of our causal sound level adjustment is to tune the audio waveform to the appropriate loudness.
It adjusts the energy of the waveform on half the length of the STFT frame at a time with the 
WebRTC auto gain control (AGC) algorithm.
Specifically, within the half-frame length of the waveform, we query the private gain experience table to obtain the gain according to the calculated amplitude, and then apply it to the waveform.
The gained waveform is then passed through the STFT to obtain the complex spectrum.

\subsection{Restoration module}

Excessive suppression of the damaged speech signal by the enhancement module may lead to the speech signal not being restored correctly.
To avoid this issue, we first employ the restoration module for speech distortion restoration, narrowband BWE and primary denoising and dereverberation.

Previously available restoration models usually generate speech waveform based on the mel-domain \cite{kong2020hifi}, which is borrowed from Text-to-Speech (TTS) task.
Nevertheless, the poor high-frequency representation of the generative model and the inadequate utilization of phase information by the mel-domain generative model render them both inappropriate for the complex scenario of this challenge. It is well known that phase recovery is helpful for speech enhancement. In this paper, we propose a complex spectrum mapping-based GAN as the restoration module by leveraging recent advances in speech enhancement and speech synthesis.

The generator of CSM-GAN follows the ``encoder-sequence modeling-decoder'' architecture, which takes the complex spectrum as the input and obtains the corresponding restoration results. As Fig.\ref{fig:ganarch} shows, the encoder contains a convolution layer followed by the dense block, and 3 convolution layers are stacked after that. The decoder comprises the corresponding transposed convolution layers and transposed convolution-dense layers. The kernel of the convolution layer is set to (2, 3) in the time and frequency axis, and the stride is (1, 2). 
Between the encoder and decoder, there are stacked temporal convolutional network blocks for temporal modeling. Skip connections are added to avoid gradient vanishing.
To reduce the number of parameters and computational effort, the fullband complex spectrum is divided into 3 subbands, and we then concatenate them in the channel dimension and hand them over to the generator finally.

Regarding the discriminators, multi-resolution frequency discriminators \cite{liu2021basis} and our proposed multi-band discriminators are adopted together.
Multi-resolution frequency discriminators are composed of stacked convolutional blocks, which are used to capture spectral structures of different frequency resolutions. 
The magnitude spectrum and its logarithmic spectrum are concatenated as the input. Each discriminator is composed of 7 2D convolution layers with a kernel size of $(3, 3)$ and a stride of $(1, 1)$ or $(2, 2)$. Weight normalization and LeakyReLU are applied sequentially after each convolution layer except the last one. For the multi-band discriminators, the network architecture is the same as the multi-resolution frequency discriminator, while a band spectrum is replaced as the input. With the multi-band discriminators, the problem of a large dynamic range in different subbands is overcome.

With CSM-GAN, we can fully utilize the phase information and efficiently tackle the high-frequency components of speech.

The training loss comprises a combination of components: a term for reconstruction loss, a term for adversarial loss, and a term for feature match loss. The reconstruction loss is made up of a multi-resolution fullband and subband short-time Fourier transform (STFT) loss. 

To achieve the multi-resolution STFT loss, we minimize the spectral convergence loss~{\cite{arik2018fast}}, along with the L1 distance in the logarithmic magnitude spectral domain, while utilizing various FFT analysis parameters, which can be written as:
\begin{equation}
\label{eq:mrstftloss}
\mathcal{L}_{s}(X) = \sum_{r}{\left(||log(X_r)-log(\widehat{X}_r)||_1 + \mathcal{L}_{sc}(X)\right)},
\end{equation}
where $X_r$ and $\widehat{X}_r$ are the spectrum of the clean speech and the predicted waveform with FFT-point of $r$. The spectral convergence loss function can be written as:
\begin{equation}
    \mathcal{L}_{sc}(X)=\frac{||X_r-\widehat{X}_r||_F}{||\widehat{X}_r||_F}.
\end{equation}

The loss functions for the fullband and subband cases are denoted by $\mathcal{L}_{s}(S)$ and $\mathcal{L}_{s}(S^{sub})$, respectively. Here, $S$ represents the magnitude spectrum of the complete signal $s$, whereas $S^{sub}$ corresponds to the subband signals $s^{sub}$ obtained by decomposing the signal using pseudo-quadrature mirror filters (PQMF).

To train the generator and discriminator, we use LS-GAN~{\cite{mao2017least}} adversarial loss. This ensures that the generator is able to deceive the discriminator during training. Additionally, the LS-GAN helps the discriminator distinguish between clean samples (labeled as 1) and samples estimated by the restoration module (labeled as 0). The generator $G_S$ and discriminator $D_S$ loss functions are given by:

\begin{gather}
\label{adv}
 \mathcal{L}_{adv} = \mathbb{E}\left[(1-D_S(\hat{s}))^2\right],\\
\mathcal{L}_{D_S} = \mathbb{E}\left[(D_S(s)-1)^2 +(D_S(\hat{s}))^2\right], 
\end{gather}

Moreover, to reduce the L1 distance between the feature maps of the discriminator for genuine and synthesized audio, a feature matching loss is computed, as presented in~{\cite{kumar2019melgan}}. This approach has proven to be successful in previous research. 

 \begin{equation}
\label{eq:feat}
\mathcal{L}_{feat} = \mathbb{E}\left[ \frac{1}{L}\sum_{l=0}^{L-1}{|D_S^{l}(s) - D_S^{l}(\hat{s})|} \right],
\end{equation}
which $L$ denotes the number of the discriminator's layer.

The generator's total loss is a combination of the aforementioned loss components, weighted appropriately:
 \begin{equation}
\label{eq:feat}
\mathcal{L}_{G_S} = \mathcal{L}_{s}(S) + \mathcal{L}_{s}(S^{sub})  + \lambda_{adv}\cdot\mathcal{L}_{adv} + \lambda_{feat}\cdot\mathcal{L}_{feat},
\end{equation}
where $\lambda_{adv}$ and $\lambda_{feat}$ are set to 1 and 20, respectivelty.

\subsection{Enhancement module}
There may still exist residual noise and artifacts in the output of the restoration module.
We apply the enhancement module in the second stage to eliminate these residual noises and artifacts to further improve the quality of the speech signal.

For maintaining the performance and reducing the computational efforts, we conduct the fullband-wideband parallel processing in the enhancement module. Note that this paper tends to provide a framework for solving complex speech impairment problems, where the networks are all replaceable.
More specifically, we divide the fullband complex spectrum into two groups of features: the complex spectrum of the wideband speech and 32 equivalent rectangular bandwidth (ERB) bands containing fullband information by spliting bands.
Subsequently, the wideband TaylorEnhancer \cite{li2022general} (TaEr) and fullband masking-based UNet (FBM UNet) \cite{schroter2022deepfilternet2} are introduced to handle the wideband complex spectrum and ERB bands in parallel, respectively. They are trained from scratch.
Taer is an all-neural denoising framework to mimic the
behavior of Taylor’s series and can be modeled as the superimposition of the 0th-order and high-order polynomials, where the former only concerns the magnitude recovery and the latter polynomials are tasked with complex-residual estimation.
TaEr has superior wideband noise suppression capability and focuses on wideband speech enhancement, while the FBM UNet provides the advantage of low complexity for fullband processing.
The outputs of the two sub-networks are then integrated into the enhanced fullband complex spectrum by the band-merge operation.

The loss function is defined as:
\begin{equation}
\label{eq:taer_loss}
\mathcal{L}(X) = \lambda_{cplx} \times \mathcal{L}_{cplx}(X) + \lambda_{mag} \times \mathcal{L}_{mag}(X),
\end{equation}
where $\lambda_{cplx}$ and $\lambda_{mag}$ denote the weights of the complex loss function and the magnitude loss function, respectively. The complex spectrum loss function and the magnitude spectrum loss function can be defined as:
\begin{equation}
\label{eq:cplx_loss}
\mathcal{L}_{cplx}(X) = \left|\left||X|^{0.3}\frac{X}{|X|}-|\widehat{X}|^{0.3}\frac{\widehat{X}}{|\widehat{X}|}\right|\right|_2^2,
\end{equation}
\begin{equation}
\label{eq:taer_loss}
\mathcal{L}(X) = |||X|^{0.3}-|\widehat{X}^{0.3}|||_2^2,
\end{equation}

\section{Experiments}
\label{experiments}

\begin{figure*}[ht]
\centering
    \includegraphics[width=0.7\linewidth]{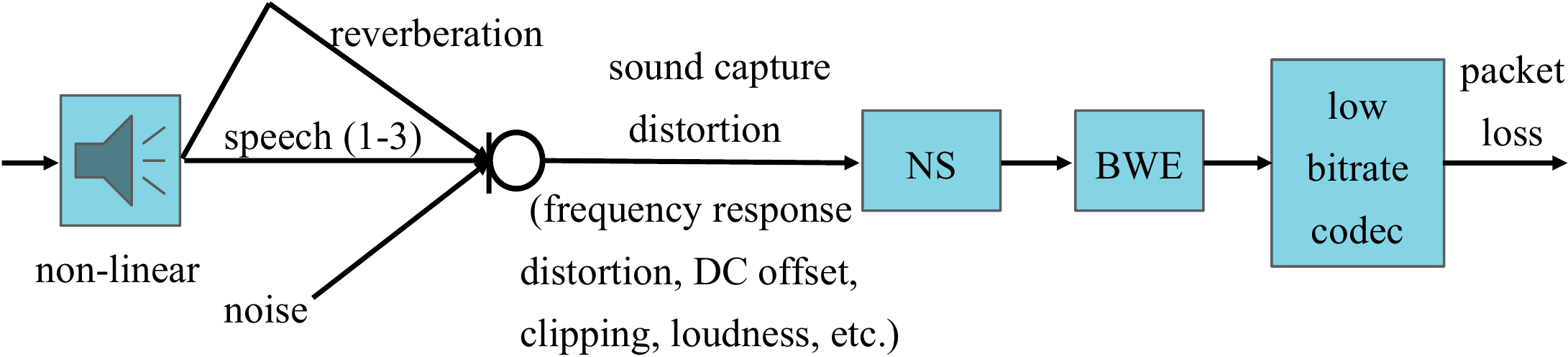}
	\caption{The pipeline of data simulation, where ``NS" refers to noise suppression and ``DC" indicates direct current.}
	\label{fig:datasim}
 \vspace{-0.3cm}
\end{figure*}

\subsection{Datasets}
We randomly selected subsets from the DNS Challenge corpus \cite{dubey2022icassp} and our internal dataset with different sampling rates as our clean set and the noise set. For convenience, all the clean and noise data were resampled to 48kHz.
The room impulse responses (RIRs) were generated based on the image method.
We subjectively analyzed the problematic audio from SSI challenge devset and simulated a 1500-hour dataset according to the proportion of various specific cases of issues including coloration, discontinuity, loudness, background noise and reverberation, etc. 

The training data simulation procedure was shown in Fig.~\ref{fig:datasim}. Specifically, the clean input proportionally with non-linear distortions was firstly mixed with noise and reverberation to generate the noisy-reverberant data. And then, to simulate the various cutoff frequencies and distortion types of the receiving microphone, the noisy-reverberant data is processed by a low-pass filter with different cutoff frequencies ranging from 1kHz to 24kHz and applied with various receiver distortions such as spectral leakage, clipping, half-wave rectification, etc. After that, the received data was processed by our private noise suppressor (NS) and blind bandwidth extension (BWE) module. Finally, several open-source codecs (AAC~{\cite{wolters2003closer}}, OPUS~{\cite{valin2012rfc}}, etc.) with different bit-rates and packet loss rates were adopted to the BWE output to simulate RTC network transmission.

The test set is provided by the organizer, which is a blind set of 500 devices/environments, which have an approximately uniform distribution for the impairment areas including noisiness, coloration, discontinuity, loudness and reverberation. 

\subsection{Experimental setup}
We applied the Hanning window with a 20 ms window length and a 10 ms frame shift. All utterances are segmented into 4 seconds. The models are trained with a maximum step of 20000000 with AdamW optimizer. The learning rate is 2e-4 and the batch size is set to 16.

\section{Results and Analysis}
\label{results-and-analysis}

In this section, we evaluate the proposed system across the objective and subjective evaluation. For the objective evaluation, DNSMOS~{\cite{reddy2021dnsmos}} and NISQA~{\cite{mittag2021nisqa}} are chosen to evaluated the performance of the systems, where DNSMOS is a non-intrusive perceptual objective speech quality metric to evaluate noise suppression, and NISQA is a non-intrusive objective speech quality assessment metric to evaluate speech quality and naturalness including noisiness, discontinuity, loudness and coloration. Subjective listening test includes two tests. 
The first evaluation metric is based on P.835, measures SIG, BAK, and OVRL while the second evaluation is based on an extension of P.804 (listening phase) and P.863.2 (Annex A) and relies on crowdsourcing and the P.808 toolkit developed by the organizers.

\subsection{Ablation study}
The ablation study spans the following three aspects: First, we verify the superiority of the complex spectrum mapping-based GAN namely CSM-GAN over the GAN in the time domain namely TD-GAN which takes SEA-Net as the generator and has the same discriminator with CSM-GAN.  Then, the necessity of the ``restoration and enhancement'' framework namley Gesper is verified, and the ``enhancement and restoration'' architecture is called as ER-Net.
Table~{\ref{table:obj}} and Fig.~{\ref{fig:spec}} show the performance of these methods. According to the table, several observations can be discovered. Firstly, CSM-GAN shows better performance 
than TD-GAN, although TD-GAN has higher computational cost. This is because that high-frequency components which correspond to the fine structures of speech signals are hard to modeling for the time-domain generator. Secondly, Gesper outperforms the CSM-GAN, indicating that the following enhancement module is necessary to remove artifacts generated by CSM-GAN and further suppress noise and reverberation components. Finally, it can be found that the speech quality processed by ER-Net is poor compared with Gesper or even CSM-GAN, demonstrating that applying noise reduction in the first stage will severely damage the speech signal in the complexity acoustic cases, which results in the following regeneration model being unable to recover these components. The above ablation experiments show that the validity and reasonableness of the proposed ``restoration and enhancement'' framework.

\begin{table}[htbp]
\centering
\caption{Comparisons of DNSMOS new and NISQA using different methods, The best results are boldfaced. {Noi.}, {Dis.}, Col. and {Loud.} indicate noisiness, discontinuity, coloration and loudness, respectively.}
\label{table:obj}
\scalebox{0.84}{
\begin{tabular}{c|ccc|ccccc}
\hline
\multirow{2}{*}{Methods}  &  \multicolumn{3}{c|}{DNSMOS} & \multicolumn{5}{c}{NISQA}\\
\cline{2-9}     
 & {SIG} & {BAK} & {OVR} & {MOS} & {Noi.} & {Dis.} & {Col.} &{Loud.}\\ 
 \hline
noisy & 2.89 & 3.45 & 2.46 & 2.34 & 3.11 & 3.39 & 2.80 & 2.87\\
CSM-GAN & 3.44 & 4.03 & 3.14 & 3.74 & 4.03 & 4.19 & 3.67 & 3.94\\
TD-GAN & 3.27 & 3.99 & 2.98 & 3.24 & 3.79 & 3.78 & 3.28 & 3.66\\
ER-Net & 3.33 & 3.98 & 3.02 & 3.33 & 3.65 & 3.81 & 3.33 & 3.62\\
Gesper & \textbf{3.45} & \textbf{4.12} & \textbf{3.20} & \textbf{3.97} & \textbf{4.33} & \textbf{4.28} & \textbf{3.79} & \textbf{4.09}\\
\hline
\end{tabular}}
\vspace{-0.3cm}
\end{table}
\begin{table}[htbp]
    \begin{center}
    \caption{Subjective evaluation results based on ITU-T P.835 on the SSI Challenge blind test set.
    } 
    \vspace{-0.1cm}
    \label{table:1}
    \scalebox{0.91}{
    \begin{tabular}{cccc}
    \toprule
        \multirow{2}{*}{Methods}  &  \multicolumn{3}{c}{ITU-T P.835 MOS} \\
    \cmidrule(lr){2-4}
    &  Overall & Signal & Background   \\
    \midrule
    Noisy &  2.824 & 3.147 & 3.453  \\
    Hitiot & 3.089 & 3.312 & 4.074 \\
    Gesper &  3.350 & 3.581 & 4.208  \\

    \bottomrule
    \end{tabular}}
    \end{center}
    \vspace{-0.5cm}
\end{table}

\begin{table}[!htbp]
    \vspace{-0.3cm}
    \begin{center}
    \caption{Part of the subjective evaluation results based on P.804 on the SSI Challenge blind test set.
    } 
    \label{table:2}
    \scalebox{0.84}{
    \begin{tabular}{ccccc}
    \toprule
        \multirow{2}{*}{Methods}  &  \multicolumn{4}{c}{P.804 MOS} \\
    \cmidrule(lr){2-5}
    &  Coloration & Discontinuity & Loudness & Reverberation   \\
    \midrule
    Noisy    & 3.029 & 4.061 & 2.992 & 3.852 \\
    Hitiot   & 3.248 & 4.005 & 3.916 & 4.477  \\
    Gesper    & 3.598 & 4.201 & 4.109 & 4.316  \\

    \bottomrule
    \end{tabular}}
    \end{center}
    \vspace{-0.7cm}
\end{table}

\begin{figure}[!htbp]
	\centering

        \subfigure[The unprocessed speech.]{
\includegraphics[width=0.45\linewidth]{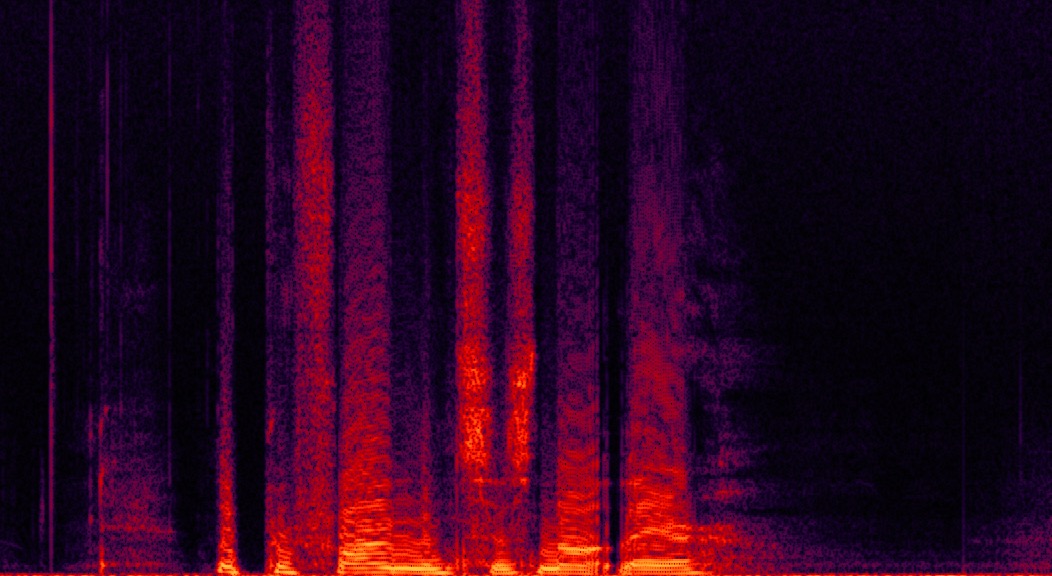} 
}
\subfigure[The speech processed by TD-GAN.]{
\includegraphics[width=0.45\linewidth]{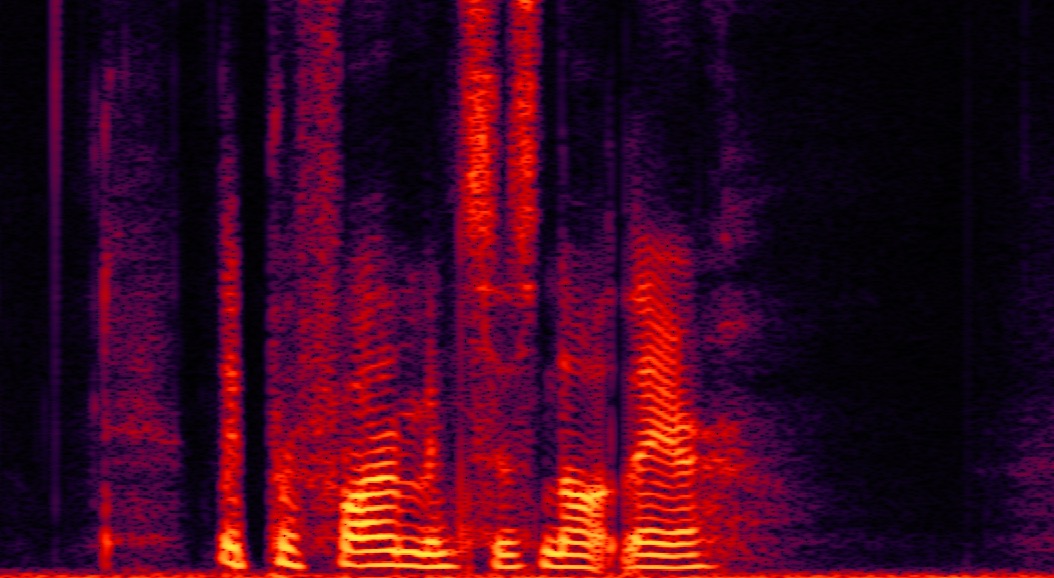} 
}
\subfigure[The speech processed by CSM-GAN.]{
\includegraphics[width=0.45\linewidth]{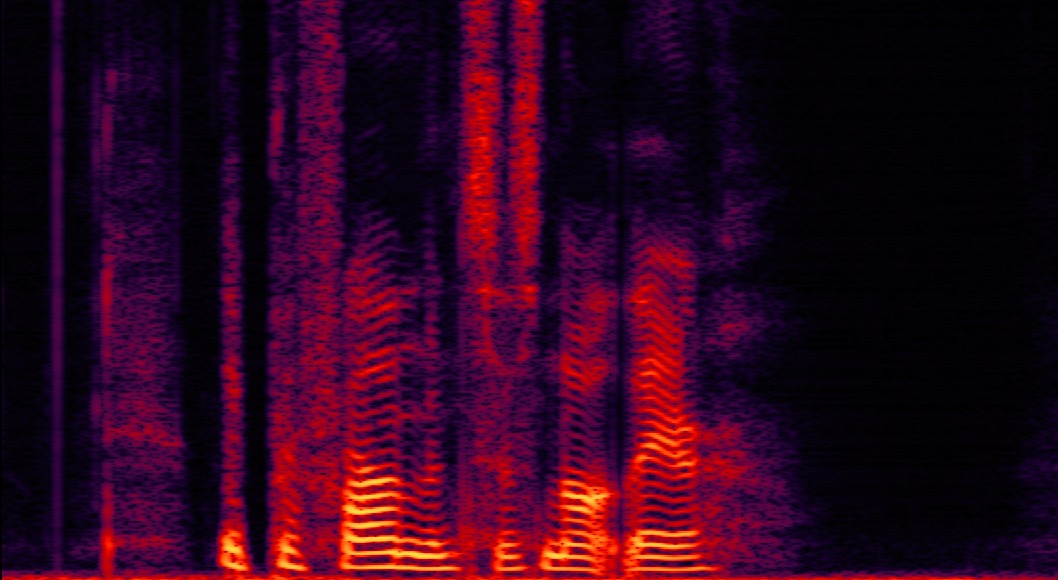} 
}
\subfigure[The speech processed by Gesper.]{
\includegraphics[width=0.45\linewidth]{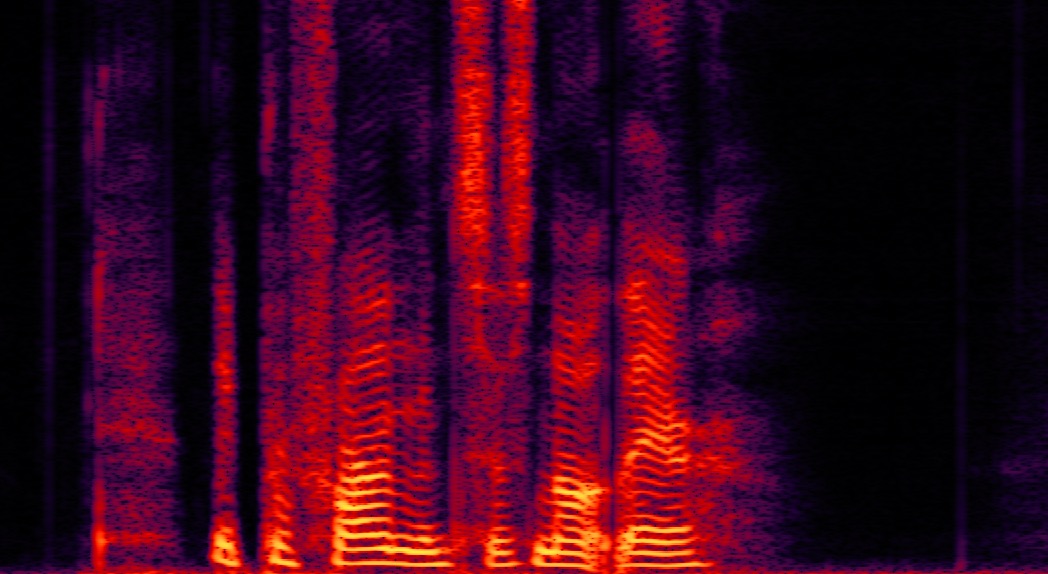} 
}
\caption{Spectrograms of results.}
	\label{fig:spec}
    \vspace{-0.3cm}
\end{figure}

\subsection{Evaluation on the SSI Challenge blind test set}
Table \ref{table:obj} reports the performance of Gesper in terms of DNSMOS and NISQA. Compared to the baselines, the proposed system achieves significant improvement in terms of all metrics consistently, 0.74 DNSMOS and 1.63 NISQA gains are obtained, respectively.
Table \ref{table:1} and Table \ref{table:2} show partial results of a multi-dimensional subjective test in term of subjective evaluation on the SSI Challenge blind test set. 
And the proposed system yields a significant improvement in all metrics relative to the noisy signals and other submissions.
This indicates that our proposed system efficiently alleviates the difficulties of noise, coloration, discontinuity, loudness and reverberation, which play a vital role in speech signal quality.

\subsection{Parameter number and real-time factor}
Moreover, we counted the number of parameters and the real-time factor (RTF). The proposed model has a total parameter number of 12.1 M, and its RTF on an Intel Core i5 Quadcore CPU (clocked at 2.4 GHz) with single thread is 0.37.

\section{Conclusions}
\label{conclusion}

This paper introduces our submission to the ICASSP 2023 SSI Challenge.
Our proposed two-stage framework achieves impressive results in addressing the challenges of noise, coloration, discontinuity, loudness and reverberation that reduce the speech quality.
The proposed real-time system is ranked first place in tracks 1 and 2 of the ICASSP 2023 SSI Challenge.


\vfill\pagebreak
\bibliographystyle{IEEEtran}
\bibliography{refs}

\end{document}